\documentclass{aastex63}

\usepackage{gensymb}
\usepackage{amssymb}
\usepackage{hyperref}
\usepackage[caption=false]{subfig}
\usepackage [autostyle, english = american]{csquotes}
\MakeOuterQuote{"}

\begin{document}

\title{Improvements to the Search for Cosmic Dawn Using the Long Wavelength Array}

\correspondingauthor{Christopher DiLullo}
\email{christopher.dilullo@nasa.gov}

\author[0000-0001-5944-9118]{Christopher DiLullo}
\affiliation{University of New Mexico \\ 
210 Yale Blvd. NE \\ Albuquerque, NM 87131, USA}
\affiliation{Lab for Observational Cosmology, Code 665, NASA Goddard Space Flight Center \\
8800 Greenbelt Rd. \\ Greenbelt, MD 20771, USA}

\author[0000-0003-1407-0141]{Jayce Dowell}
\affiliation{University of New Mexico \\ 
210 Yale Blvd. NE \\ Albuquerque, NM 87131, USA}

\author[0000-0001-6495-7731]{Gregory B. Taylor}
\affiliation{University of New Mexico \\ 
210 Yale Blvd. NE \\ Albuquerque, NM 87131, USA}

\begin{abstract}
We present recent improvements to the search for the global Cosmic Dawn signature using the 
Long Wavelength Array station located on the Sevilleta National Wildlife Refuge in New Mexico, USA 
(LWA--SV). These improvements are both in the methodology of the experiment and the hardware of 
the station. An improved observing strategy along with more sophisticated temperature calibration 
and foreground modelling schemes have led to improved residual RMS limits. A large improvement over 
previous work using LWA--SV is the use of a novel achromatic beamforming technique which has been developed for LWA--SV. 
We present results from an observing campaign which contains 29 days of observations between March 
$10^{\rm{th}}$, 2021 and April $10^{\rm{th}}$ 2021. The reported residual RMS limits are 6 times above the amplitude 
of the potential signal reported by the Experiment to Detect the Global EoR Signature (EDGES) collaboration.

\end{abstract}


\section{Introduction} \label{sec:intro}

Detecting the formation of the first stars, known as Cosmic Dawn, and probing the subsequent 
Epoch of Reionization (EoR) remain some of the biggest goals of observational cosmology. 
The efforts to detect signals from these times in cosmic history focus either on 
direct imaging of the first luminous sources in the infrared, a major science goal of the 
upcoming James Webb Space Telescope \citep{JWST}, or detecting the redshifted 21 cm signal 
from neutral hydrogen present during these periods \citep{furlanetto2006review, morales2010, pritchard2012}. 
The redshifted 21 cm signal offers the only ground-based approach to studying Cosmic Dawn and 
the EoR since Earth's atmosphere is opaque in the relevant infrared band.

The 21 cm signal provides an especially strong probe which traces the evolution of neutral 
hydrogen throughout the Epoch of Reionization. Measuring the brightness of the signal across frequency 
would give insights into how astrophysical conditions evolve as the first luminous sources begin to dominate
the Universe \citep{cohen2017}. The 21 cm signal is generally broken into two observable signals: 
a global signal which is detectable over large angles on the sky and a spatially inhomogeneous signal 
which corresponds to the growth of HII regions during the EoR. The global Cosmic Dawn signal is expected
to manifest as a small absorption feature in the average spectrum of the sky below 100 MHz. Experiments to 
detect the 21 cm signal focus on detecting either the 
global signal by measuring the average spectrum of the sky and accurately modelling and removing 
foreground signals \citep{sokolowski2015, singh2018, price2018, mckinley2020} or the inhomogeneous signal by 
measuring the three dimensional k-space power spectrum whose structure arises from the formation 
of ionized regions during the Epoch of Reionization \citep{parsons2010, deboer2017,mertens2020,yoshiura2021}.
Novel interferometric techniques have also been developed for future use in detecting the global signal 
\citep{mckinley2012,vedantham2015,mckinley2020}.

The potential detection of a global absorption signal reported by the Experiment to Detect
the Global EoR Signature (EDGES) collaboration \citep{bowman2018} has sparked much interest and
debate due to its unexpected shape and amplitude \citep{hills2018, bradley2019, singh2019, sims2019}.
If validated, the EDGES absorption signal would imply that current cosmological models fail to accurately 
predict the astrophysical processes which govern the 21 cm physics of neutral hydrogen throughout Cosmic 
Dawn and the Epoch of Reionization. This hints to either new physics in the form of interactions between 
dark matter and baryons \citep{barkana2018, munoz2018, berlin2018} or the presence of a previously unaccounted 
for radio background \citep{feng2018, mirocha2018, dowell2018}. 

The implications of the unexpected nature of the absorption signal reported by EDGES warrant independent 
validation using a different instrument. The Long Wavelength Array \citep[LWA;][]{taylor2012,cranmer2017} located
in New Mexico, USA is a radio telescope whose frequency coverage allows it to possibly
validate the EDGES signal. Initial work using the LWA station located on the Sevilleta National Wildlife Refuge, NM, USA 
(LWA--SV) has been previously presented \citep{dilullo2020}. In that work, a novel beamforming approach was 
presented which allows for increased sensitivity compared to single dipole experiments, improved foreground 
minimization, and \textit{in situ} astronomical calibration. It was concluded that the dependence of 
beam shape\footnote{Throughout this work, the phrase "beam shape" is used to represent the combination of both the forward 
gain and the angular structure of the main lobe of the beam.}
with frequency is a major challenge and that a custom beamforming framework would have to be developed for LWA--SV which would
allow for achromatic beamforming.

The work detailed here describes recent improvements to the methods described in \citet{dilullo2020} which have improved 
the residual RMS limits by an average factor of 3.  These include hardware improvements to LWA--SV, changes to the 
observational strategy, improvements to the data analysis, and the successful adoption of achromatic beamforming. 
The paper is structured as follows: Section \ref{sec:improvements} details the recent changes to LWA--SV and to the 
observational strategy; Section \ref{sec:beamforming} details the achromatic beamforming framework, discusses tests to 
compare the sensitivity of a standard LWA--SV beam to that of an achromatic beam, and presents a software package which 
can be used to simulate the beam pattern, both standard and custom, for a general antenna array; Section \ref{sec:results}
will present results from a recent observing campaign which employs all the recent improvements; and Section 
\ref{sec:conclusion} discusses these results and looks towards future work.

\section{Recent Improvements} \label{sec:improvements}

The improvements to the methods described in \citet{dilullo2020} fall into two categories: methodological 
improvements to the observing strategy and data analysis and improvements to the hardware and software 
of LWA--SV. 

\subsection{Observing Strategy and Data Analysis} \label{sec:observing}

The observing strategy laid out  in \citet{dilullo2020} focuses on the use of simultaneous observations of a cold 
region on the sky (RA $9^{\rm h} \ 38 ^{\rm m} \ 40.56 ^{\rm s}$,  Dec $+ 30^{\degree} \ 49' \ 1.4 ''$; J2000), referred 
to as the Science Field, and Virgo A (RA $12^{\rm h} \ 30 ^{\rm m} \ 49.42 ^{\rm s}$, Dec $+ 12^{\degree} \ 23' \ 28 ''$;
J2000). Simultaneous observations of Virgo A allowed for astronomical calibration which would help account for any time 
dependent variation in the system. However, this did not account for the differences between the shapes of the science 
and the calibration beams as they were pointed towards different locations on the sky. In addition to the instantaneous 
differences in beam shapes, the different declinations of the Science Field and Virgo A meant that the science and 
calibration beams would trace different paths across the sky. These two effects caused inaccuracies in the 
calibration of the science beam.

A new Science Field center pointing (RA: $11^{\rm h} \ 00 ^{\rm m} \ 49.42 ^{\rm s}$, Dec: $+ 12^{\degree} \ 23' \ 28 ''$;
J2000)  has been identified. The choice of coordinates still places the Science Field in a large cold region on the sky, 
but guarantees that the science and calibration beams trace out the same arc on the sky. The observations of the Science 
Field and Virgo A are also no longer done simultaneously, but rather during times when the targets align in terms of 
position on the sky. This ensures that the derived calibration better accounts for the changing forward gain of the beam 
throughout a single observation.

\begin{figure}
    \centering
    \includegraphics[width=\textwidth]{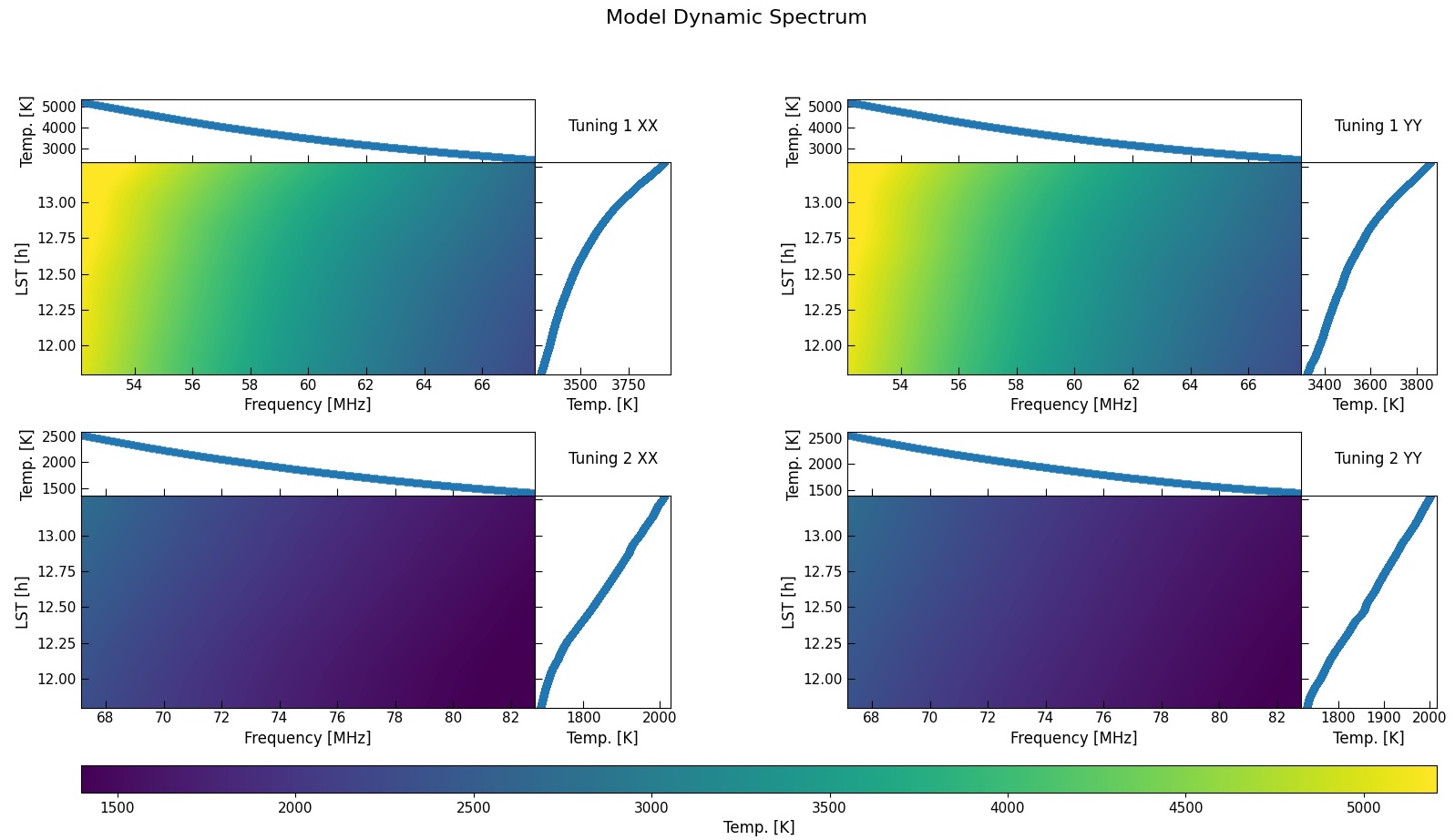}
    \caption{Model dynamic spectrum of the Virgo A beam. These are generated by simulating the beam
    pattern at every $5^\textrm{th}$ pointing for every $5^\textrm{th}$ frequency and 
    convolving these simulations with the Global Sky Model at 7.5 s resolution to yield temperature 
    spectra. These spectra are then interpolated at the frequency and time resolution of an observation. 
    The median spectrum and drift are shown at the top and right of each plot, respectively.}
    \label{fig:model}
\end{figure}

The data analysis methodology has been improved through the adoption of fully time dependent temperature calibration and 
Bayesian foreground modelling. Time dependent temperature calibration is achieved by modelling the beam pattern of LWA--SV 
for every $5^\textrm{th}$ pointing throughout the run at every $5^\textrm{th}$ observed frequency in the band and 
convolving the models with the Global Sky Model \citep[GSM;][]{deOliveira2008}. This creates a set of model spectra 
between the relevant Local Sidereal Time (LST) ranges which can be linearly interpolated in both time and frequency at the 
same temporal and spectral resolution of an observation to yield a model dynamic spectrum of the Virgo A field which is 
used to calibrate the raw data. 
It should be noted that the choice to use every $5^\textrm{th}$ pointing and frequency is purely arbitrary and was chosen 
to reduce required compute time. These models can be generated at higher temporal and spectral resolution before
interpolation, but this becomes quite computationally expensive and is not expected to significantly change the 
results since the sky changes smoothly over time and frequency and this smooth nature is accurately captured through
interpolation. An example of the model dynamic spectrum of the Virgo A beam, interpolated at 40 ms time resolution and 
 19.1 kHz frequency resolution, is shown in Figure \ref{fig:model}.

Bayesian foreground modelling allows for the posterior distributions of the model parameters to be explored to search for
correlations between foreground model parameters and 21 cm model parameters. Correlations between the foreground and 21 cm 
models would be non-physical and would imply issues with our foreground models. We use the Python package 
\textit{emcee} \citep{foreman2013} which uses the Affine Invariant Markov chain Monte Carlo (MCMC) Ensemble 
Sampler \citep{goodman2010} to maximize the likelihood function, which essentially measures the probability of a 
dataset given a certain set of model parameters. Bayes' theorem relates the posterior probability distribution, 
$\mathcal{P}(\mathbf{\Theta} | \mathbf{D}, \mathcal{H})$, of a set of parameters 
$\mathbf{\Theta}$ given the data $\mathbf{D}$ and some model $\mathcal{H}$ to the likelihood 
$\mathcal{L}(\mathbf{D | \mathbf{\Theta}, \mathcal{H}})$ via
\begin{equation}
  \mathcal{P}(\mathbf{\Theta} | \mathbf{D}, \mathcal{H}) = \frac{\mathcal{L}(\mathbf{D | \mathbf{\Theta}, \mathcal{H}}) \Pi(\mathbf{\Theta}|\mathcal{H})}{\mathcal{Z}(\mathbf{D} | \mathcal{H})} ,
\end{equation}
where $\Pi(\mathbf{\Theta}|\mathcal{H})$ is the prior distribution that encodes information about the model parameters and 
$\mathcal{Z}(\mathbf{D} | \mathcal{H})$ is the evidence, which is the integral of the likelihood function over the prior 
domain. We follow \citet{harker2012} and \citet{bernardi2016} by assuming Gaussian measurement noise in a single frequency
channel which allows us to write the likelihood $\mathcal{L}_{i}$ of observing the sky temperature $T_{sky}(\nu_{i})$ at 
frequency $\nu_{i}$ as
\begin{equation} \label{eq: likelihood}
    \mathcal{L}_{i} (T_{sky}(\nu_{i}) | \mathbf{\Theta}) = \frac{1}{\sqrt{2\pi\sigma^{2}(\nu_{i})}} e^{-\frac{(T_{sky}(\nu_{i}) - T_{m}(\nu_{i}, \mathbf{\Theta}))^2}{2 \sigma^{2}(\nu_{i})}} ,
\end{equation}
where $T_m$ is the sky model and $\sigma$ is the standard deviation of the instrumental noise is a given frequency 
channel given by
\begin{equation}
    \sigma(\nu_{i}) = \frac{T_{sky}(\nu_{i})}{\sqrt{\Delta \nu \Delta t}} ,
\end{equation}
where $\Delta \nu$ and $\Delta t$ are the channel bandwidth and integration time, respectively.

Given Equation \ref{eq: likelihood}, the log-likelihood of the entire temperature spectrum can be represented as
\begin{equation}
    \rm{ln} \mathcal{L}(\mathbf{T}_{sky} | \mathbf{\Theta}) = \sum_{i=1}^{N} \rm{ln} \mathcal{L}_{i}(T_{sky}(\nu_{i}) | \mathbf{\Theta}) ,
\end{equation}
where we have switched to the log-likelihood due to computational efficiency. The sky model is a superposition of the 
foreground contribution and the 21 cm signal, written as
\begin{equation} \label{eq: foreground}
    T_{m} (\nu_i) = T_{f}(\nu_i) + T_{21cm}(\nu_i) ,
\end{equation}
but for the work presented in this paper we do not attempt to fit the 21 cm signal as our residual root mean square (RMS) 
is still too large to meaningfully fit a 21 cm signal (See Section \ref{sec:results}). We use a slightly modified foreground 
polynomial than that used in \citet{bernardi2016}; however, it is still a $N^{\rm{th}}$ order log-polynomial given by
\begin{equation} \label{eq: log_poly}
    T_{f}(\nu_i) = \sum_{n=0}^{N} p_n \left[\textrm{log}_{10} \left(\frac{\nu_i}{\nu_0}\right) \right]^n ,
\end{equation}
where $\nu_0$ is a reference frequency that is set to the center of the band. We choose a uniform 
prior for each model parameter, $p_n$, which assigns equal probability to all real valued solutions.

\subsection{Upgrades to LWA--SV} \label{sec:lwasv}
There have been a few upgrades to the hardware of LWA--SV which should help in the search for Cosmic Dawn.
The observational bandwidth of LWA--SV has been improved to 20 MHz per tuning per beam. This new bandwidth allows for 
uninterrupted frequency coverage in the range $50 - 85$ MHz with center tuning choices of 60 and 75 MHz. Accounting for 
bandpass rolloff, we select the inner 80\% of the band which yields a usable frequency range of $52 - 83$ MHz for the 
output spectra. This frequency range should allow for the capture of the lower edge of the EDGES signal through the center 
frequency of 78 MHz. This is a large improvement over the spectra presented in \citet{dilullo2020}, which had incomplete 
frequency coverage in the smaller range of $\approx 63 - 79$ MHz.

Along with the upgraded bandwidth capabilities of LWA--SV, a new weather station was also deployed at the station which 
records the outside temperature. We have previously seen that temperature variations inside the electronics 
shelter can affect performance of the system, namely the analog receiver (ARX) boards, and so we also expect outside 
temperature variations to affect the performance of the front end electronics (FEE) on each dipole. Outside temperature data 
from the new weather station can be combined with the temperature data from the ARX boards to allow for a 2-dimensional 
fit which better captures gain fluctuations than the previous simple 1-dimensional fit with ARX temperature. However, 
we have found this effect to be so small in magnitude that it is difficult to accurately remove this trend from the data 
which is dominated by the changing sky. 

The temperature change, both outside and in the electronics shelter, over a single 
1.5 hour observation is less than 1\% compared to the change in sky brightness due to the rising Galactic plane. The temperature 
dependence of the FEE gain has been reported by \cite{hicks2012} to be in the range 
0.0042 -- 0.0054 dB/\degree C between 20 -- 100 MHz, which corresponds to a linear 
gain difference of $\approx 0.001 / \degree \rm{C}$. This small gain variation combined with the small temperature changes over 
the course of a single observation implies the FEE gain variation is negligible. The temperature dependence of the ARX gain 
has not been directly measured, but we estimate it to be of similar magnitude to the FEE temperature dependence.

Perhaps as we push our residuals below 1 K, this correction will become necessary, but for now this approach does not seem to improve data 
quality. Another improvement to note is the installation of second temperature sensor for the HVAC controller within the
electronics shelter. This has led to much more stable temperatures within the shelter and hence more stable ARX temperatures. 
We now see typical temperature variations on the order of $\approx 2\degree$ F.

\section{Achromatic Beamforming} \label{sec:beamforming}
A major challenge that all experiments attempting to detect Cosmic Dawn and the EoR face is the 
frequency dependence of the antenna response. This "chromaticity" of the receiving element 
produces spectral structure which can obscure any potential cosmological signal. The array nature 
of LWA--SV helps as beamforming allows for better rejection of foreground contributions by 
focusing the main lobe in a region of the sky away from the Galactic plane; however, this introduces 
a second chromaticity factor which can obscure the signal. The beam pattern of an antenna array is not 
only a function of the antenna gain pattern, which is chromatic, but also has intrinsic chromaticty. 
This second chromaticity factor can be combated through a custom beamforming framework which modifies 
the required complex beamforming coefficients in order to tailor the array response to be more constant 
across the observed frequency band. However, this still does not completely account for the chromaticty 
of the gain pattern of the individual antennas within the array.

In \citet{dilullo2020}, a framework for custom beamforming at a single frequency was presented. 
The beam pattern of LWA--SV can be shaped by modifying the amplitude of the complex gains of each 
dipole in the array depending on the frequency of interest and the desired pointing center on the sky. 
Model custom beam patterns were presented and driftscan observations of Cygnus A (RA $19^{\rm h} \ 
59 ^{\rm m} \ 428.36 ^{\rm s}$,  Dec $+ 40^{\degree} \ 44' \ 2.1 ''$; J2000) at transit confirmed 
that the beam full width at half maximum (FWHM) could be predictably shaped using the framework. 
However, the beam was only properly tuned for a single frequency. It was concluded that in order 
to achieve near-achromatic beamforming, modifications would need to be made to the beamformer 
pipelines of LWA--SV.

\subsection{Implementation on LWA--SV}
The beamformer pipelines are part of the digital backend of LWA--SV, known as the Advanced 
Digital Processor \citep[ADP;][]{cranmer2017, price2017, dowell2020Memo}. The ADP beamformer 
pipelines have been modified in order to allow for 
pre-computed custom complex gains to be read in and overwrite the complex gains which would 
normally be computed by the Monitor and Control System at LWA--SV. Therefore, for an observation 
which consists of some number of pointings which track a source on the sky, the necessary 
custom complex gains can be computed for every observed frequency at each pointing. This yields 
a set of custom complex gains which keep the beam pattern mostly constant in both frequency and 
time as the beam tracks the source across the sky. However, the beam pattern can still not be 
said to be completely constant as the response pattern of the individual dipoles has both angular 
and frequency dependence.

The major trade-off for the ability to shape the beam pattern of an antenna array is a loss in 
sensitivity. The sensitivity of an array is given by its System Equivalent Flux Density (SEFD)
\begin{equation}
    SEFD = \frac{k_B T_{sys}}{A_e} 10^{26} \ \rm{Jy},
\end{equation}
where $k_B$ is the Boltsmann constant, $T_{sys}$ is the system temperature in K, and $A_e$ is the 
effective area of the array in $\rm{m}^2$. The SEFD is an estimate of the flux density of a source which would 
double the system temperature, which means that a lower SEFD corresponds to a higher sensitivity. 
$A_e$ is difficult to accurately measure for an antenna array, but
the simple estimate of 
\begin{equation}
    A_e \approx N_{dip} \times A_{ant},
\end{equation}
where $N_{dip}$ is the number of antennas in the array and $A_{ant}$ is the effective 
area of the antenna, is an idealized estimate. However, effects such as mutual coupling 
of individual antennas can drastically change the effective area of the array as a 
whole \citep{ellingson2007}.

Compared to standard beamforming at LWA--SV where every antenna is equally weighted, adjusting 
the amplitudes of the complex gains to shape the beam effectively reduces the effective area 
of the entire array. Therefore, it is expected for the SEFD of the array to increase as more 
dipoles are down weighted to maintain beam shape with frequency. The SEFD of the array can be 
estimated by observing a bright source, such as Cygnus A, and comparing the observed power 
when the source is in and out of the beam. The SEFD can be estimated with
\begin{equation}
    SEFD = \frac{S_{\nu}}{\frac{P_{on}}{P_{off}} - 1},
\end{equation}
where $P_{on}$ is the observed power when the beam is pointed on source, $P_{off}$ is the 
observed power when the beam is pointed off source, and $S_\nu$ is the flux density of the source 
at the measured frequency, $\nu$. This is typically done by pointing a beam at the transit position 
of a bright source and letting the source drift into and out of the beam. Driftscans can also be used 
to measure the FWHM of the beam main lobe. The major limitation of driftscan observations is that 
only a single dimension of the beam can be probed throughout the observation, so measuring the beam 
shape in both dimensions on the sky requires either multiple beams or multiple days of observation where the 
beam pointing center is below, on, and above the source transit position so the source can probe different 
parts of the beam main lobe. 

A method to avoid the need for multiple days of observations is to continuously repoint the
beam in a basket weave pattern. This approach is based on methods used by higher frequency 
arrays such as the Very Large Array. The idea is to make two cuts along the source in different 
directions on the sky. This allows for the 2-dimensional beam shape to be estimated at various 
points in a single observation. However, at low frequencies ionospheric scintillation on 
$\sim 10 \ \rm{s}$ timescales can make things difficult since the observed power at any 
position is a combination of the beam pattern and the local ionospheric conditions. Thus, 
we modify the above approach to interleave the off-source pointings with on-source pointings. 
These on-source pointings serve as an ionospheric references which bracket each off-source 
pointing and can be interpolated between to determine the amount of ionospheric scintillation.

\begin{figure}
	\centering
	\subfloat[Standard beamforming]{%
	\includegraphics[width=0.95\linewidth]{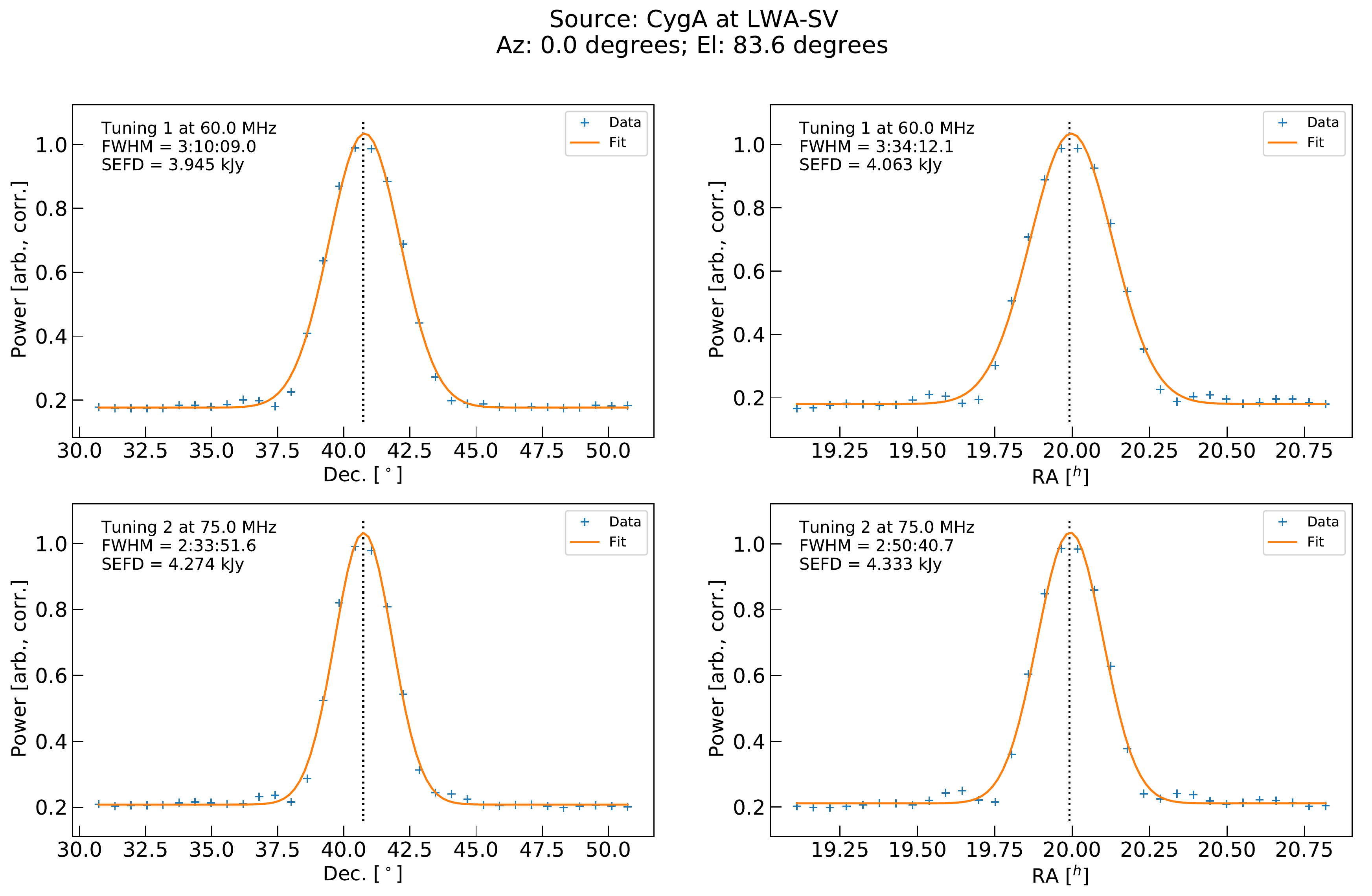}%
	} \\
	\subfloat[Achromatic beamforming]{%
	\includegraphics[width=0.95\linewidth]{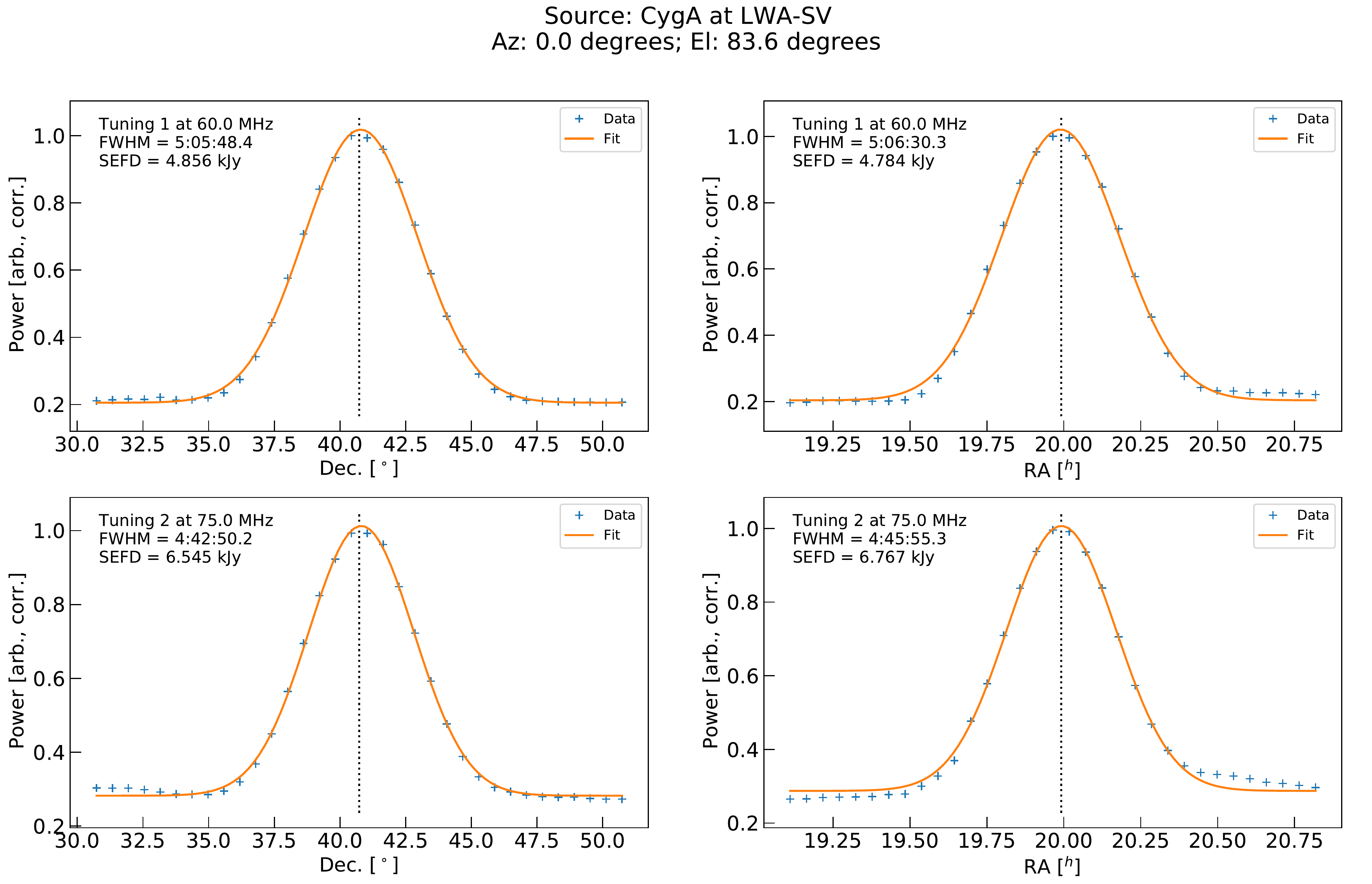}%
	}
	\caption{Interferometric basket weave results from observations of Cygnus A.
    The cuts along declination and right ascension are shown in each figure in the 
    left and right columns, respectively. The lower and upper tunings are shown in each
    figure in the top and bottom rows, respectively. The achromatic beam shows less 
    variation in FWHM across frequency in both directions on the sky.}
    \label{fig:weaves}
\end{figure}

The modified basket weave approach is utilized on a weekly basis at LWA--SV in order to 
monitor any changes in the station's SEFD. These weekly observations utilize standard beamforming 
on Cygnus A, so Cygnus A was a natural choice to use when testing the sensitivity of the station 
with an achromatic beam. A circular beam with FWHM of $4\degree$ was chosen for a basket weave 
observation of Cygnus A; however, estimating the SEFD of LWA--SV turned out to be challenging as 
the larger beam size meant that more power from the Galactic plane was being picked up by the beam. 
This generally smoothed out the contribution of Cygnus A to the total observed power and resulted 
in unrealistic SEFD estimates. To avoid this, we used the beam-dipole observing mode which LWA--SV 
offers to observe Cygnus A interferometrically using the modified basket weave to determine the 
phase centers. Beam-dipole mode correlates the beamformed output with the signal from a single antenna 
in the array. We chose to correlate the beamformed output with the outrigger antenna which is located 
$\sim 300 \ \rm{m}$ to the West of the station center. This two-element interferometer produces a 
fringe pattern on the sky which is a function of the observed frequency, the baseline geometry, and 
the geometric mean of the beam patterns of the station beam and the outrigger dipole. These 
interferometric basket weave observations resolve out the diffuse structure from the Galactic plane 
and help isolate the contribution from Cygnus A. Results from observations using standard and achromatic 
beamforming can be seen in Figure \ref{fig:weaves}. The final SEFD estimates are averages between the two 
values derived from the RA and Dec slices, respectively. The SEFD and FWHM estimates  for beam-dipole mode 
observations using standard and achromatic beamforming are presented in Table \ref{table:Weaves}. 
It is clear that the observations using achromatic beams suffer a loss in sensitivity; however, the 
loss in sensitivity is less than a factor of 2.

\begin{deluxetable}{ccc}
\tablewidth{0pt}
\tablecaption{\label{table:Weaves} Beam-Dipole Mode Basket Weave Results}
\tablehead{\colhead{Frequency} & \colhead{Standard Beamforming} & \colhead{Achromatic Beamforming}}
\startdata
                & SEFD: $4.0 \ \rm{kJy}$ & SEFD: $4.8 \ \rm{kJy}$ \\
        60 MHz  &      & \\
                & FWHM: $3^{\degree} 22' 10.6''$ & FWHM: $5^{\degree} 06' 09.3''$ \\
\hline \\
                & SEFD: $4.3 \ \rm{kJy}$ & SEFD: $6.7 \ \rm{kJy}$ \\
        75 MHz  &      & \\
                & FWHM: $2^{\degree} 42' 16.1''$ & FWHM: $4^{\degree} 44' 22.7''$ \\
\enddata
\end{deluxetable}

\subsection{Beam Simulator: A Python Package to Simulate Array Beam Patterns}
The achromatic beamforming work presented above might be of interest to a broader community that 
is interested in either simply simulating the beam pattern of a general antenna array or developing 
similar custom frameworks for an antenna array. A Python package, aptly named \textit{Beam Simulator}, has 
been developed to help simulate the beam pattern of a general antenna array. It can represent the 
elements of an antenna array down from a single cable up to the entire array and is also compatible 
with output files from the Numerical Electromagentics Code (NEC) which is used to numerically model 
the gain pattern of antennas. This allows for the simulated beam pattern to encapsulate the gain pattern of the
individual array elements and not just account for the geometry of the array. \textit{Beam Simulator} 
also allows for custom beams to be simulated using the same framework that has been developed for 
LWA--SV. There are convenience functions built in to easily represent a LWA station, so this 
should be of particular interest to any institutions that plan on hosting a future LWA Swarm 
station \citep{Taylor2019}. However, it should also be of interest to the more general community 
interested in representing the beam pattern of any phased array. The package can be found 
and downloaded on GitHub at \url{https://github.com/cdilullo/beam_simulator}.

\section{Recent Observing Campaign and Results} \label{sec:results}
An observing campaign was carried out between March $10^{\rm{th}}, \ 2021$ and 
April $10^{\rm{th}}, \ 2021$ in order to build a sufficiently large data set. A single 
observation consists of an achromatic beam with a main lobe FWHM of $4{\degree}$ 
which tracks the Science Field (see Section \ref{sec:observing}) between the local 
sidereal time (LST) range $10.28 - 11.78 \ {\rm h}$ and a second achromatic beam 
with a similar main lobe FWHM which tracks Virgo A between the LST range $11.78 - 13.28 \ {\rm h}$.
We use the spectrometer observing mode offered by LWA--SV which yields dynamic spectra with 40 ms 
time resolution and 1024 frequency channels each having 19.1 kHz of bandwidth. The individual 
observations from each day are flagged for RFI in both time and frequency by applying 
a median smoothing window to the median drift and spectrum, respectively, to create smooth 
models of each and then flagging times or frequencies which deviate from these models. 
This allows for the RFI flagging to be controlled through 4 simple parameters: the smoothing 
window sizes and the threshold values in both time and frequency. It was found that the 
parameters which led to good RFI detection were a time smoothing window of $30 \ \rm{s}$, 
a frequency smoothing window of $250 \ \rm{kHz}$, and time and frequency thresholds of $3\textrm{-}\sigma$ and 
$5\textrm{-}\sigma$, respectively, where $\sigma$ is the standard deviation of the data after the smoothed model has been 
divided out. These parameters led to relatively aggressive flagging for a single observation, $\approx 20\%$ 
of data for each day, but this was intentional to try and capture low level RFI.

A common source of RFI in the relevant frequency band is ionospheric reflections from terrestrial 
digital television (DTV) signals which manifest as flares with a characteristic 6 MHz bandwidth. 
DTV channels $2\rm{-}6$ all lie within our observed band and can be seen quite frequently. Due to the short 
timescales of the bursts and the relatively small bandwidth that is affected, these DTV flares are not 
captured in the above RFI flagging method which is better suited at capturing long duration narrow-band RFI 
or short duration wide-band RFI. To better capture these events, we search along the pilot 
tone frequencies for each channel for times where the pilot tone is $2\textrm{-}\sigma$ above the median 
value and flag the entire 6 MHz sub-band corresponding to the DTV channel at these times.

\begin{figure}
    \centering
    \includegraphics[width=\textwidth]{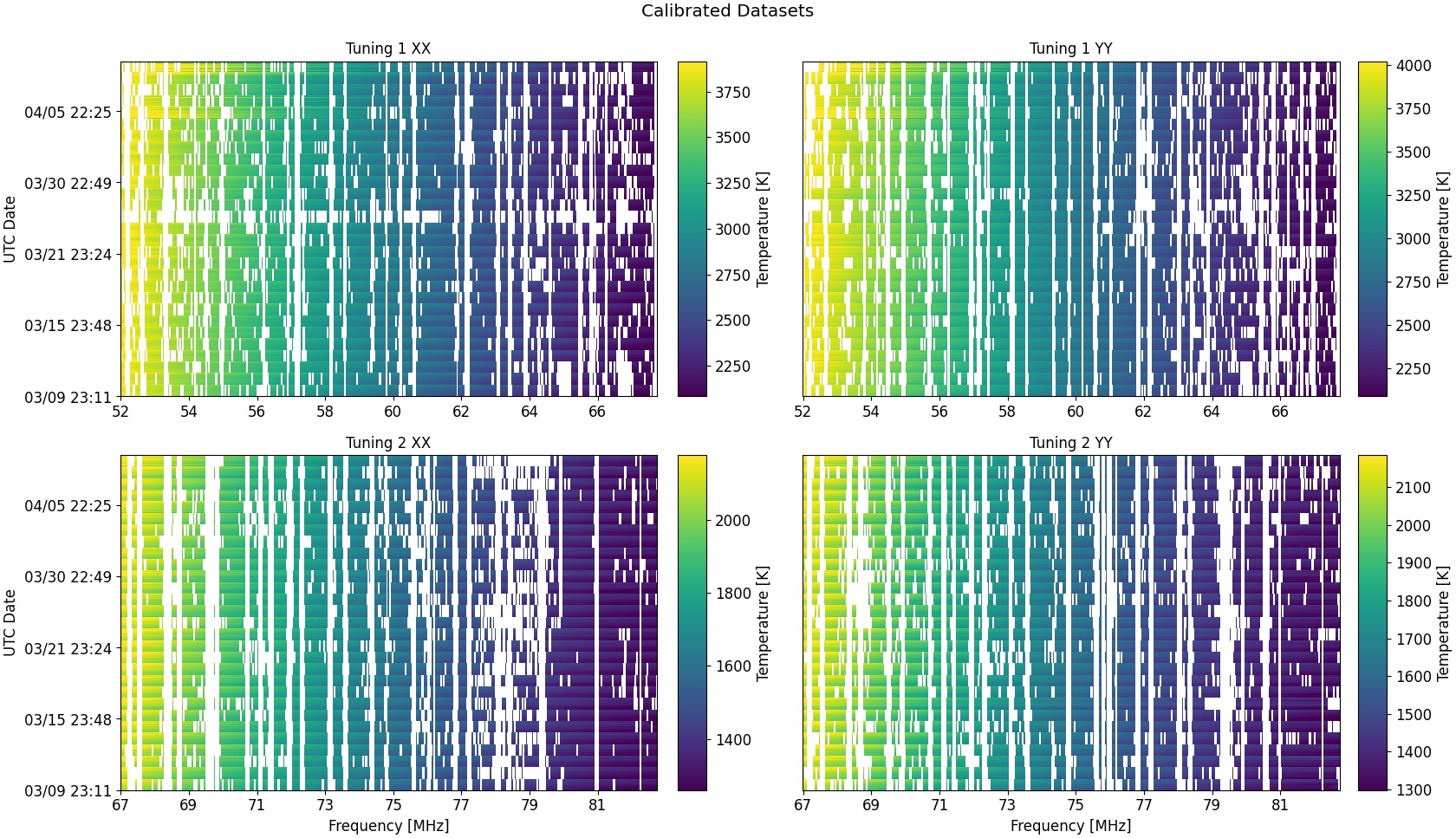}
    \caption{Calibrated datasets for the entire observing campaign at 30 second integration
    time. The individual 1.5 hour observations can be seen as a repeating pattern along the 
    time axis. The full dataset consists of 29 days of data taken between March 
    $10^{\rm{th}}, \ 2021$ and April $10^{\rm{th}}, \ 2021$.}
    \label{fig:cal_data}
\end{figure}

After each day was flagged for RFI, we performed a moving median window smoothing operation on the 
data in time in order to smooth out variations due to ionospheric scintillation. Ionospheric 
scintillation causes random variations in the brightness of our targets with time, which obscure 
the temperature calibration. We chose a window size of 5 minutes in order to properly smooth out
scintillation. The individual flagged and smoothed datasets were then temperature calibrated by 
multiplying a model temperature dynamic spectrum of Virgo A (see Section \ref{sec:observing}) 
with the ratio between the raw Science Field and Virgo A data. The calibrated Science Field data were 
then integrated in time into 30 second bins before being combined into a single dataset which contains 
the entire observing campaign. An extra RFI flagging step was applied which flags any channel that has 
been flagged for more than 75\% of the entire campaign. The fully flagged, calibrated, and integrated 
dataset for the entire observing campaign is shown in Figure \ref{fig:cal_data}.

\begin{figure}
    \centering
    \includegraphics[width=\textwidth]{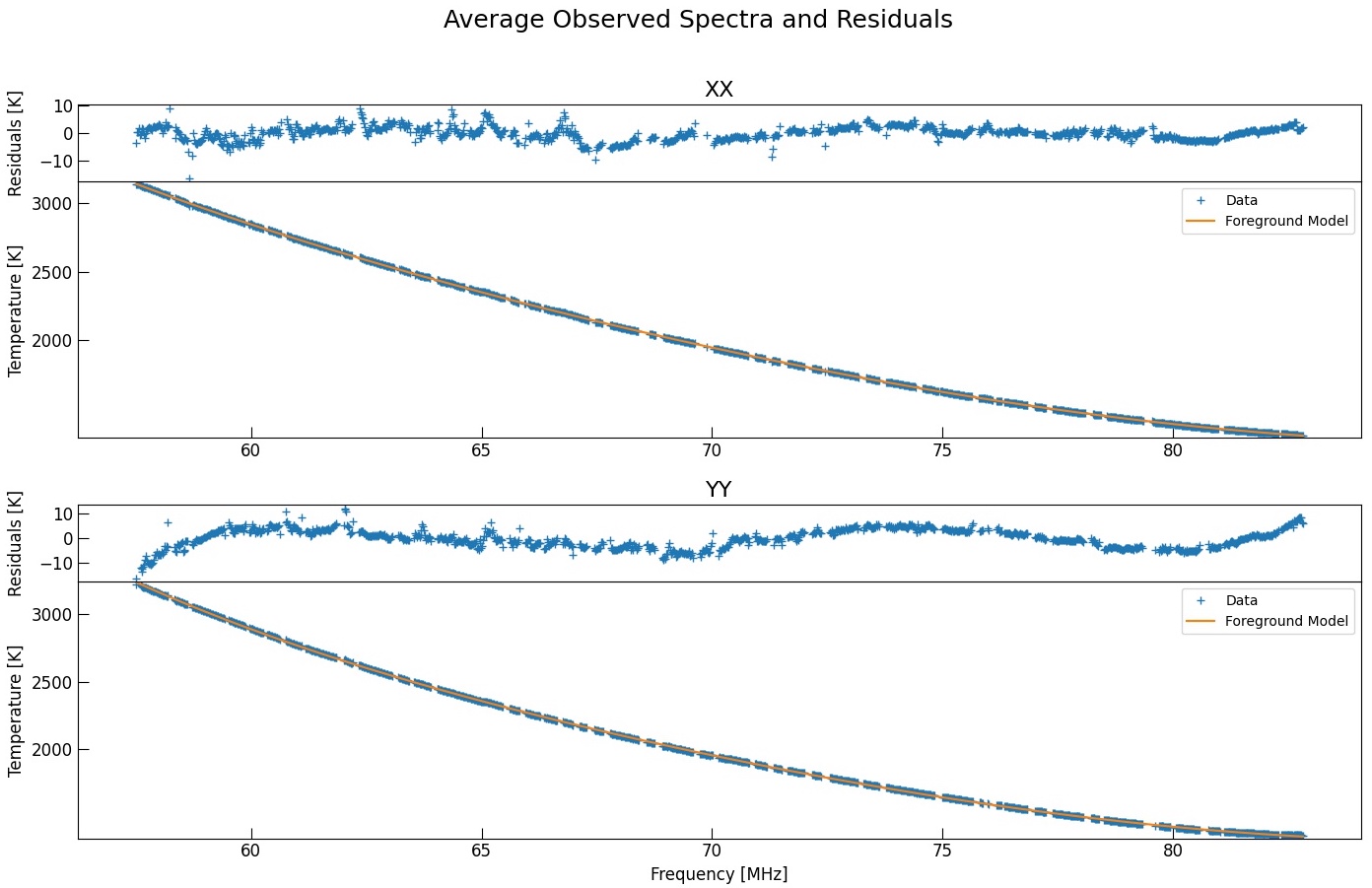}
    \caption{Results of the MCMC fit.}
    \label{fig:mcmc}
\end{figure}

\begin{figure}
    \centering
    \includegraphics[width=\textwidth]{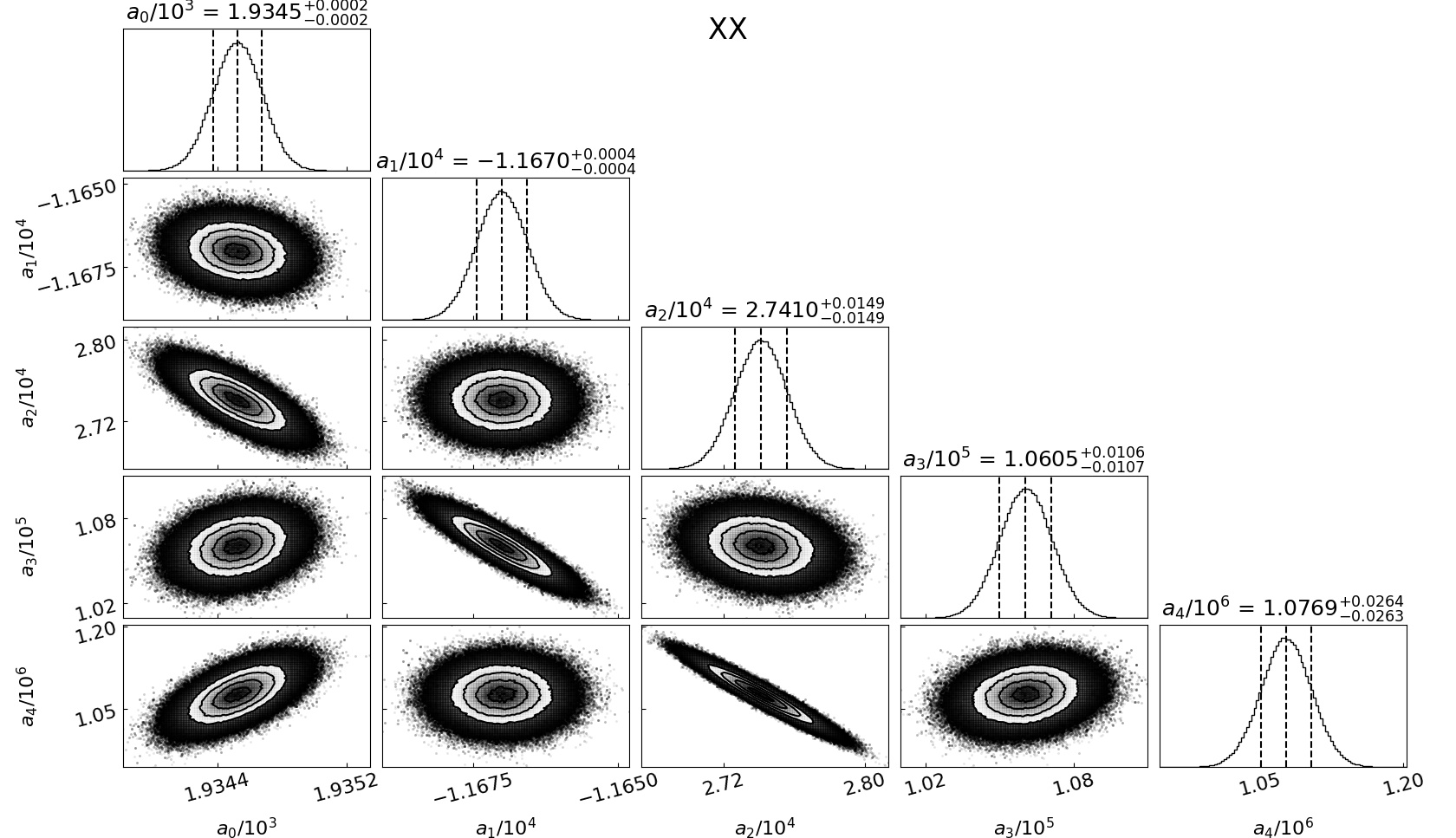}
    \caption{Posterior distributions for foreground model parameters for Tuning 1 XX.
    The 1-dimensional posteriors show the median value along with the $16^{\rm{th}}$ and $84^{\rm{th}}$ 
    percentile bounds. This plot was made with the Python package \textit{corner} \citep{foreman2016}.}
    \label{fig:posteriors}
\end{figure}

We generated a measured spectrum by employing a bootstrapping algorithm to generate samples from the full 
calibrated dataset. This bootstrapping algorithm ensured that the output spectrum was the most "typical" 
and was not influenced by random features, such as unflagged RFI, which were present in only some of the 
integrated spectra. In order to generate the samples, we first fit a smooth bandpass model to each integrated 
spectrum which was then divided out to create a copy of the data which highlighted the unsmooth features in each 
integrated spectrum. We then computed the standard deviation across frequency for each of these residual spectra 
and computed the first and third quartiles, $Q1$ and $Q3$, for the distribution of standard deviations. We flagged 
outlier spectra whose standard deviation was above the standard outlier threshold given by
\begin{equation}
    \sigma \geq 1.5 \times (Q3 - Q1) .
\end{equation}
Out of a total of 5046 integrated spectra for the entire campaign, 405 outlier spectra were identified for Tuning 1 XX, 
274 for Tuning 1 YY, 163 for Tuning 2 XX, and 301 for Tuning 2 YY.
The remaining non-outlier integrated spectra for each tuning and polarization were then bootstrap sampled. 
The bootstrap sampling randomly samples the set of non-outlier spectra and returns a set of 
the same size. An average spectrum is then computed using the resampled set of spectra.
This was repeated 10,000 times in order to generate a set of average spectra for each tuning and 
polarization. The average of these 10,000 spectra was taken to be the "typical" spectrum. The two average spectra for a given 
polarization were then combined to yield a single average spectrum which spans the entire bandwidth of 52--83 MHz. 
A foreground model was then fit using the Bayesian modelling framework described in Section \ref{sec:observing}.
We chose to start 100 walkers around the zero vector and run a chain which was 25,000 steps long. It was observed that the model 
struggled to properly fit the data across the entire frequency band, with frequencies below $\approx 57.5$ MHz having the poorest 
fit, so the lower edge of the band below 57.5 MHz was discarded before the fitting. The average bootstrapped spectrum after trimming,
along with the fit foreground model and residuals, for each polarization is shown in Figure \ref{fig:mcmc}. 
Figure \ref{fig:posteriors} shows a typical plot of the posterior distributions of the model parameters for the XX polarization. 
A similar plot is generated for the YY polarization.

\begin{figure}
    \centering
    \includegraphics[width=\textwidth]{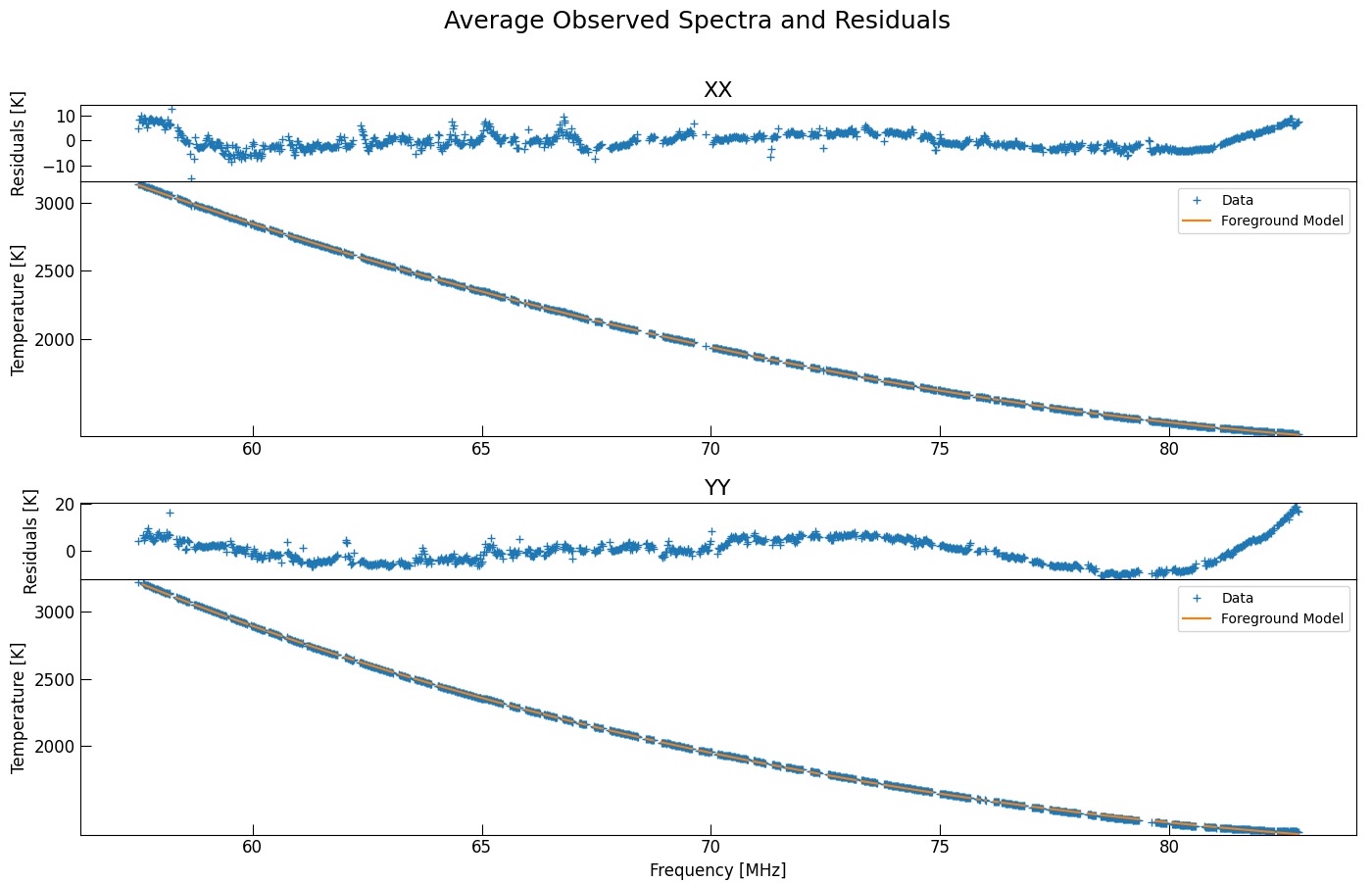}
    \caption{Results of the maximally smooth function foreground 
            model fit. }
    \label{fig:msf}
\end{figure}

We also investigate the performance of maximally smooth functions \citep[MSFs;][]{rao2015, rao2017} as 
foreground models. The Python package \textit{maxsmooth} \citep{bevins2021} is used to fit the same 
$5 \textrm{th}$ order log-polynomial foreground model (See Equation \ref{eq: log_poly}). We find that the 
MSF model performs well, but there appears to be higher order spectral structure which is not captured. 
This is not surprising as MSFs are designed to be maximally smooth so that they preserve the non-smooth features 
of the data in residuals. The observed spectrum along with the MSF model and residuals are shown in Figure \ref{fig:msf}. 
The residual RMS across frequency for each polarization using both modelling methods is summarized in Table \ref{table:RMS}.

\iffalse
\begin{deluxetable}{ccccc}
\tablewidth{0pt}
\tablecaption{\label{table:RMS} Residual RMS Across Frequency}
\tablehead{\colhead{Foreground Modelling Method} & \colhead{Tuning 1 XX} & \colhead{Tuning 1 YY} & 
\colhead{Tuning 2 XX} & \colhead{Tuning 2 YY}}
\startdata
MCMC                       & 3.50 K & 3.32 K & 1.36 K & 1.24 K \\
Maximally Smooth Function  & 5.05 K & 5.99 K & 5.26 K & 6.25 K
\enddata
\end{deluxetable}
\fi

\begin{deluxetable}{ccc}
\tablewidth{0pt}
\tablecaption{\label{table:RMS} Residual RMS Across Frequency}
\tablehead{\colhead{Foreground Modelling Method} & \colhead{XX} & \colhead{YY}}
\startdata
MCMC                       & 2.47 K & 3.81 K \\
Maximally Smooth Function  & 3.29 K & 5.26 K 
\enddata
\end{deluxetable}

\section{Discussion and Conclusion} \label{sec:conclusion}
The residual RMS limits reported in Table \ref{table:RMS} are encouraging as they are beginning to approach 
the sub-Kelvin level that is required for a proper validation of the reported EDGES signal. The desired goal 
is to attain a residual RMS of $\approx 50 \rm{mK}$ so that the EDGES signal could be validated at the 
$10\textrm{-}\sigma$ level. However, pushing the residual RMS to the sub-Kelvin level is not expected to be simple.

We believe that the largest systematic that is limiting our residual RMS is the discrepancy between the beam 
model used in the temperature calibration and the true beam pattern of LWA--SV on the sky. The model beam pattern 
predicts a very smooth response which is unlikely to be realistic. We tested these differences by taking an observation 
from a single day of the observing campaign and fitting a smooth model to the data. Many spectral and 
temporal structures remained after the smooth model was removed from the data which implies that 
there are underlying low level non-smooth spectral structures present in the data. These structures are 
most likely related to the sidelobe pattern of the array which is difficult to measure and quantify.

There are also discrepancies between the model used for the gain pattern of a single LWA dipole and the true gain 
pattern. The model is directly used in the simulation of the beamformed pattern of the station as a whole and so 
is also present in the temperature calibration. The model is based on Numerical Electromagnetics Code (NEC) simulations 
of the LWA dipole, but directly measuring it is extremely difficult at these low frequencies. There is an ongoing 
collaboration with the External Calibrator for Hydrogen Observatories (ECHO) team at Arizona State University whose 
goal is to measure the gain pattern of the LWA dipole through use of a drone which transmits a tone of known power 
at various altitudes and azimuths \citep{chang2015, jacobs2016}.

Another concern is the presence of low level RFI which is difficult to detect and flag via the median spectrum and 
drift method described in Section \ref{sec:results}. Any unflagged RFI will significantly increase the residual RMS, 
especially as we approach the sub-Kelvin level. The bootstrapping algorithm described in Section \ref{sec:results} 
should help reduce the contribution of random low level RFI which is only present on some days of the observing campaign, 
but more systematic low level RFI which is present throughout the entire observing campaign at the same frequencies will 
still be problematic. As we decrease our residual RMS limits, we will need to better understand the RFI environment of LWA--SV, 
whether it is external or internal to the station itself.

Information about what physically might be limiting LWA--SV also can be gleaned from the spectral structure of the residuals.
There appears to be large scale ripples in the residuals of size $\approx 20 \rm{MHz}$ which could suggest some physical 
signal reflection with a wavelength of 15 m, or even half this, is present within the system. However, there is no characteristic 
length present within the LWA--SV signal chain, so the origin of these ripples is still unknown. Ripples such as these could 
possibly manifest from other effects such as mutual coupling between antennas and cross coupling between signal paths on the 
ARX boards, but these are much more difficult to isolate if they are contributing to this structure.

It is also important to acknowledge that the above work does not calibrate out additive noise from the electronics that is 
present in the data. This additive noise will arise mainly from the amplifiers present in the FEE boards on each antenna and 
the ARX boards which process the analog signal. \citet{hicks2012} has reported the noise temperature of the FEE boards to be 
in the range 255--273 K between 20--100 MHz, but the noise temperature of the ARX boards has not been directly measured.
The noise temperature of the ARX boards could be measured by injecting a noise source of known temperature into the backend 
electronics of LWA--SV. This was done in \citet{dilullo2020}, however the purpose of that work was to verify that the system 
was stable over time and so was not rigorously designed to measure the noise temperature of the ARX boards. We leave a targeted 
study of the ARX noise temperature as a function of frequency using signal injection for future work. The experience of other experiments 
\citep{rogers2012, monsalve2017, singh2018, nambissan2021} which have made similar measurements suggests this may not be easy; 
however, our sky based calibration should capture the majority of effects like this and so our noise measurements may not need to be so painstaking.

The work presented above shows encouraging improvements to the previously reported limits of LWA--SV. Future 
work will focus on better quantifying the LWA dipole gain pattern and understanding discrepancies between models 
of the station beam and the true beam pattern. This is crucial if the residual RMS is to be reduced to the level 
where the 21 cm signal can be jointly fit with the foreground model. A more robust RFI detection algorithm may have 
to be developed in order to capture any low level RFI that is present throughout the entire observing campaign.

\acknowledgments
Construction of the LWA has been supported by the Office of Naval Research under Contract N00014-07-C-0147 and by 
the AFOSR. Support for operations and continuing development of the LWA is provided by the Air Force Research 
Laboratory and the National Science Foundation under grants AST-1835400 and AGS-1708855.

\facility{LWA}

\bibliographystyle{aasjournal}
\bibliography{References.bib}

\end{document}